\documentstyle[preprint,aps]{revtex}

\newcommand{\bra}{\langle}
\newcommand{\ket}{\rangle}
\newcommand{\sixj}[6]{
     \left\{ \begin{array}{ccc}
              #1 & #2 & #3 \\
              #4 & #5 & #6
            \end{array}  \right\} }
\newcommand{\ninej}[9]{
     \left\{ \begin{array}{ccc}
              #1 & #2 & #3 \\
              #4 & #5 & #6 \\
              #7 & #8 & #9
            \end{array}  \right\} }

\begin{document}
\draft
\title{State dependent effective interaction for the hyperspherical formalism
with noncentral forces}
\author{Nir Barnea$^{1}$, Winfried Leidemann$^{2,3}$,
and Giuseppina Orlandini$^{2,3}$}
\address{$^1$The Racah Institute of Physics, The Hebrew University, 91904,
         Jerusalem, Israel\\
$^{2}$Dipartimento di Fisica, Universit\`a di Trento, I-38050 Povo, Italy;\\
$^{3}$Istituto Nazionale di Fisica Nucleare, Gruppo collegato di Trento.
}

\date{\today}
\maketitle

\begin{abstract}
The recently developed effective interaction method for the hyperspherical
harmonic formalism is extended to noncentral forces. Binding energies and radii
of three- and four-body nuclei are calculated with AV6 and AV14 NN potentials. 
Excellent results for the convergence of the expansion are found, particularly 
for the three-nucleon system. Due to the higher density the convergence rate 
is a bit slower for the alpha particle.
In comparison to central potential models there is only a very 
slight deterioration of the convergence due the tensor force, 
while other potential terms have no visible effect on the convergence. The 
obtained values for binding energy and radii also agree well with
the results in the literature obtained with other few-body techniques. 

\end{abstract}


\section{Introduction}

In the last years great progresses have been achieved in microscopic
calculations of few-body systems. Quite a few different exact approaches
have been developed. Among those the effective interaction method is
particularly interesting since it is a typical many-body rather 
than a few-body approach. In fact effective interactions are used in 
shell-model calculations of complex nuclei, where the wave functions are 
expanded in a model space consisting in  a truncated 
single particle harmonic oscillator (HO) basis. The introduction of an 
effective rather than a bare interaction is necessary in order to compensate
for the truncation of the space. A few years ago such an effective interaction 
method has also been introduced in few-body physics. So called no-core 
shell-model calculations have been carried out, where one keeps all the 
nucleons active and where one also removes the spurious center of mass motion
finally leading to HO basis functions that depend 
on the Jacobi coordinates ~\cite{Navratil98}. 

Recently we have proposed a different effective interaction method 
\cite{BLO00} using instead of the HO basis an expansion of the wave function 
in hyperspherical harmonics (HH). In the HH formalism
the Jacobi coordinates are replaced by a single length
coordinate, the hyperradius, and a set of $3A-4$ hyperangles.
The HH are the A-body generalization of the 2-body spherical harmonics,
and likewise depend only on the hyperangular (angular) coordinates in the
hyperspherical (spherical) decomposition of the A-body (2-body) system. In 
general, the wave function can be expanded in a series consisting of products 
of HH basis functions and hyperradial basis functions.  The HH basis is 
widely used in the calculation of few-body wave functions, though the
convergence is very often problematic. In order to improve
the convergence one may introduce proper  
correlation functions (see, e.g.,~\cite{Fenin 72,Krivec90,Rosati92,BLO99}).
The correlation functions, however, lead to various undesirable features in
the calculation. One not only looses the orthogonality of the basis functions 
but also has no unique way to determine 
the correlation functions. Therefore, as an alternative approach, we have 
studied the HH effective interaction (EIHH) method  to accelerate the slow 
convergence of the expansion. Employing simple central NN potential models 
as bare NN interaction for light nuclei in the mass range of $A=3-6$ we could 
show that the EIHH method leads to an extremely rapid convergence for the 
investigated observables. One important reason for the excellent convergence 
behavior is due to the fact that the effective interaction becomes state 
dependent in a natural way, i.e. the interaction does not only depend on the 
state of the two interacting particles, but also on the state of the rest 
system.  

Due to the excellent convergence the EIHH method  has proven  to be a 
very promising 
tool in few-body physics. On the other hand, in Ref.~\cite{BLO00}
only central potential models have been considered while realistic NN
interactions are much more complicated  and in principle could deteriorate
the convergence rate.  Therefore it is of
great importance to check whether the EIHH method leads to similarly good
results also for more realistic NN potentials. In this paper we investigate
the method for two noncentral NN interaction models 
(AV6, AV14 \cite{AV14}) calculating ground state energies and radii for the 
$A=3,4$ nuclei. To this end the convergence patterns are studied in detail and 
the final results are compared to results in the literature.

The paper is organized as follows. The EIHH method for noncentral forces
is described in Sec.~\ref{sec:EIHH}. Numerical results, 
including a comparison with results obtained with other few-body techniques, 
are given in Sec.~\ref{sec:num} and conclusions are drawn in Sec.~\ref{sec:con}.

\section{The HH Effective Interaction}
\label{sec:EIHH}

Before introducing the effective interaction we first give a short summary 
of the HH formalism.  
To illustrate the hyperspherical coordinates we start from the definition of 
the center-of-mass coordinate 
$\vec{ R}=\frac{1}{A}\sum_{i=1}^A \vec{ r}_i$
and the normalized reversed order $N=A-1$ Jacobi coordinates
\begin{equation} \label{Jacobi}
  \vec{\eta}_i  =  \sqrt{\frac{A-i}{A+1-i}}\Big(\vec{ r}_i 
                - \frac{1}{A-i} \sum_{j=i+1}^{A} \vec{ r}_j \Big) \,.
\end{equation} 
The Jacobi coordinate $\vec{\eta}_j$ consists of a radial 
coordinate $\eta_j$ and a pair of angular coordinates 
$\hat{\eta}_j \equiv (\theta_j,\, \phi_j)$. 
These coordinates are then transformed into the hyperangular coordinates
$\alpha_2,\ldots,\alpha_{N}$ through the relations
\begin{equation} \label{hyper_n}
  \sin \alpha_n =  \eta_n / \rho_{n} \; ,\,\,\,\,\,\,\,\,\,
\rho_{n}^2 = \rho_{n-1}^2 + \eta_n^2 = \sum_{j=1}^n \eta_j^2 \; .
\end{equation}
Note that the
hyperradial coordinate $\rho \equiv \rho_{N}$ is symmetric with respect to 
permutations of the underlying single particle coordinates.
The $3N=3(A-1)$ internal coordinates for the $A$-particle system consist of 
the hyperradial coordinate $\rho$ and the $3N-1$  ``hyperangular'' coordinates 
$\Omega_{N} \equiv \{\hat{\eta}_1,\, \hat{\eta}_2,\, \cdots,\, 
\hat{\eta}_{N}, \alpha_2,\, \alpha_3,\, \cdots,\, \alpha_{N}\} $.

With the hyperspherical coordinates one can write the Laplace operator for
$n$ Jacobi coordinates $n=1 \ldots N$,
as a sum of two terms 
\begin{equation} \label{Laplacen}
  \Delta_{n} = \frac{1}{\rho_{n}^{3n-1}}\frac{\partial}{\partial \rho_{n}} 
  {\rho_{n}^{3n-1}} \frac{\partial}{\partial \rho_{n}}
   - \frac{1}{\rho_{n}^2} \hat{K}_n^2 \; .
\end{equation}
The  hyperspherical or grand angular momentum 
operator $\hat{K}_n^2$ of the $n$ Jacobi coordinates can be expressed in 
terms of 
 the squared angular momentum associated with the nth Jacobi coordinate, 
$\hat{\ell}_n^2$, and $\hat{K}_{n-1}^2$ as follows ~\cite{Efros}
\begin{equation} \label{Kn}
 \hat{K}_n^2 = - \frac{\partial ^2}{\partial \alpha_n^2} +
 \frac{3n-6 - (3n-2) \cos (2\alpha_n)} {\sin(2\alpha_n)}
 \frac{\partial}{\partial \alpha_n} 
+\frac{1}{\cos^2 \alpha_n}\hat{K}_{n-1}^2
+ \frac{1}{\sin^2 \alpha_n} \hat{\ell}_n^2 \; , 
\end{equation}
where we define $\hat {K}_1^2 \equiv \hat{\ell}_1^2$. The 
 total  angular momentum operator associated with $n$ coordinates
is $\vec{\hat{L}}_n = \vec{\hat{L}}_{n-1} +\vec{\hat{\ell}}_n$.
The operators $\hat{K}_{n}^2$, $\hat{\ell}_{n}^2$, $\hat{K}_{n-1}^2$,
$\hat{L}_n^2$ and $\hat{L}_{n_{\mbox{\scriptsize $z$}}}$ 
commute with each other. 
The hyperspherical harmonic functions ${\cal Y}_{[K_n]}$ are the 
eigenfunctions of $\hat{K}_n^2$ with eigenvalues  $K_n(K_n + 3n - 2)$.
 The symbol 
$[K_n]$ stands for a set of quantum numbers including, e.g.,
$\ell_1,...,\ell_n$, $L_2,...,L_n$ and $K_2,...,K_n$. 
For more detailed information
on the HH functions see ~\cite{Fabre83,Efros}.

The HH functions do not possess any special
properties under particle permutation. Therefore the first step in 
applying the expansion to the A-body problem is the symmetrization
of the HH basis. In the current work we  employ two powerful algorithms
~\cite{Nir9798,Akiva94} recently developed for the construction 
of a HH basis with well defined permutational symmetry.  

Employing noncentral potentials it is usually more convenient to work in
a $jj$ coupling scheme. However, constructing symmetrized basis 
functions in this coupling scheme is more complicated, roughly by a factor 
$2^A$, in comparison to the $LS$ scheme.
 Therefore  we work in the $LS$ scheme coupling HH
basis functions with definite permutational symmetry  with  
spin-isospin basis functions into anti-symmetric many-body states with
total angular momentum $J$.

Now we turn to the question of the
effective interaction.
In general we would like to use the HH basis functions to solve the
A-body Hamiltonian
\begin{equation}\label{H_0}
  H = \sum_{i=1}^{A} \frac{\vec{p}_i^{\,2}}{2 m} + \sum_{i<j}^{A} V_{ij} \;,
\end{equation}
where $\vec p_i$ is the momentum of the $i$th particle and 
$m$ the nucleon mass, while 
\begin{equation}
  V_{ij}=V(\vec{r}_{ij}) = \sum_p V^{(p)}(r_{ij}) 
                    \hat{O}^{(p)}(\hat{r}_{ij} \;; 
                    \vec{\sigma}_i, \vec{\sigma}_j,
                    \vec{\tau}_i, \vec{\tau}_j  )
\end{equation}
is the NN interaction 
 which also depends on the spin ($\vec \sigma_i$) and
isospin ($\vec \tau_i$) operators of the two interacting particles. 
As already mentioned one usually introduces correlation functions 
in order to obtain sufficiently converging results
 ~\cite{Fenin 72,Krivec90,Rosati92,BLO99}. 
In our alternative EIHH method of Ref. ~\cite{BLO00} we use instead of the
bare NN interaction an effective interaction inside the
model space. Defining $P$ as the projection operator onto the model space and 
$Q=1-P$ as the projection onto the complementary space, the model space
Hamiltonian can be written as
\begin{equation}\label{H_P}
  H_P = P\left[ \sum_{i=1}^{A} \frac{\vec{p}_i^{\,2}}{2 m} \right] P + 
        P \left[ \sum_{i<j}^{A} V_{ij} \right]_{eff} P \;.
\end{equation}
In the HH formalism the model
space can be defined as a product of the hyperradial subspace and the 
set of HH basis functions with generalized angular momentum quantum number 
$K \leq K_{max}$.

In general the effective interaction appearing in Eq. (\ref{H_P}) is an A-body
interaction. If it is determined without any approximation, then the 
model-space Hamiltonian provides a set of eigenvalues which coincide with a 
subset of the eigenvalues of the original full-space Hamiltonian, Eq. 
(\ref{H_0}). 
 Since the effective interaction has to contain the whole A-body information
of Q-space the calculation of this 
A-body effective interaction is as difficult as finding the full-space 
solution. Because of these difficulties we introduced in Ref. ~\cite{BLO00} 
 instead of the exact A-body an approximate two-body effective interaction 
which, however, satisfies the following properties:
(i)  $V_{2\;eff} \longrightarrow V_{ij}$   
as $K_{max} \longrightarrow \infty$ and (ii)
the eigenvalues, $E_i(K_{max})$, and eigenvectors of the effective 
Hamiltonian converge to their limiting values faster than the 
eigenvalues and eigenvectors of the bare Hamiltonian. 

We expressed the HH effective interaction in terms of the matrix element of the 
 $N$th Jacobi coordinate, i.e. the 
``last'' particle pair,
\begin{equation}
  \bra \sum_{i<j}^A V_{2\;eff}(\vec{r}_{ij}) \ket = 
  \frac{A(A-1)}{2}\bra V_{2\;eff}(\vec{r}_{A,A-1}) \ket \;.
\end{equation} 
The relevant hyperspherical degrees of freedom 
associated with $V_{2\;eff}(\vec{r}_{A,A-1})$ are $\hat{\eta}_N$
and the hyperangle,
\begin{equation}
	\sin \alpha_N = \frac{r_{A,A-1}}{\sqrt{2}\rho} \;.
\end{equation}
 For the construction of $V_{2\;eff}(\vec{r}_{A,A-1})$ 
in Ref. ~\cite{BLO00} we made the following ansatz for the 
hyperspherical ``2-body'' Hamiltonian 
\begin{equation}\label{H2}
  H_2(\rho) = \frac{1}{2 m} \frac{\hat{K}_N^2}{\rho^2} 
            + V(\sqrt{2}\rho \sin \alpha_N \cdot \hat {\eta}_N) \;.
\end{equation}
It is a natural choice
since $\hat{K}_N^2$ contains the canonical kinetic energy associated
with the two-body variables $\alpha_N$ and $\hat{\eta}_N$ (see Eq. (\ref{Kn})).
Such an $H_2$ is in fact an A-body Hamiltonian.
It contains the hyperspherical part of the A-body kinetic energy and 
 thus becomes  a function of the collective coordinate $\rho$.
The hyperradial kinetic energy operator has not been included in $H_2$. 
The reason is that we can use a complete basis set for the $\rho$-space
and therefore we do not need  to
define an effective interaction for the hyperradial part. 

Due to the collective coordinate, $\rho$, in $H_2$ one has automatically  a 
confinement of the 2-body-system: for moderate values of $\rho$
the relation $0\leq r_{A,A-1}\leq\sqrt{2}\rho$ ensures localization of the
2-body wave function and for large values of $\rho$ the Hamiltonian
coincides with the bare one, since the NN interaction vanishes. Therefore 
large overlaps between the model space 
states and the eigenvectors of the 2-body problem are ensured and thus, 
different from the HO effective interaction, there is
no necessity for an additional confining potential. 

In order to diagonalize $H_2$ we have to recouple the last 2 particles into 
good spin $s$, isospin $t,t_z$ and $j$ states 
\begin{eqnarray} & &
\bra T_2 T_3 \ldots T_A T_{Az}
     \Big( ([K_{N-1}] L_{N-1} ; \ell_N)  K_N L_N ; S_2 S_3 \ldots S_A
     \Big) J J_Z |  \cr & & \hspace{3.5cm} 
     H_2 \; | T'_2 T'_3 \ldots T'_A T_{Az}
     \Big( ([K'_{N-1}] L'_{N-1}; \ell'_N) K'_N L'_N ; S'_2 S'_3 \ldots S'_A
     \Big) J J_Z \ket  
     \cr & & \hspace{1.5cm}
 =   \delta_{[K_{N-1}][K'_{N-1}]}\delta_{S_2, S'_2}
     \delta_{S_3, S'_3} \ldots   \delta_{S_{A-2},S'_{A-2}} 
     \delta_{T_2, T'_2} \ldots   \delta_{T_{A-2},T'_{A-2}} 
     \cr & & \hspace{1.5cm} \times
    \sum_{s s'} \sum_{j\; J_{A-2}} (-1)^{2S_{A-2}+S_A+S'_A}
    \sqrt{(2S_{A-1}+1)(2S'_{A-1}+1)(2s+1)(2s'+1)} 
     \cr & & \hspace{2.5cm} \times
    \sixj{S_{A-2}}{\frac{1}{2}}{S_{A-1}}{\frac{1}{2}}{S_{A}}{s}
    \sixj{S_{A-2}}{\frac{1}{2}}{S'_{A-1}}{\frac{1}{2}}{S'_{A}}{s'} 
     \cr & & \hspace{2.5cm} \times
    \sqrt{(2L_N+1)(2S_A+1)(2J_{A-2}+1)(2j+1)}
    \ninej{L_{N-1}}{\ell_N}{L_N}{S_{A-2}}{s}{S_A}{J_{A-2}}{j}{J} 
     \cr & & \hspace{2.5cm} \times
    \sqrt{(2L'_N+1)(2S'_A+1)(2J_{A-2}+1)(2j+1)}
    \ninej{L_{N-1}}{\ell'_N}{L'_N}{S_{A-2}}{s'}{S'_A}{J_{A-2}}{j}{J} 
     \cr & & \hspace{1.5cm} \times
    \sum_{t t'} (-1)^{2T_{A-2}+T_A+T'_A}
    \sqrt{(2T_{A-1}+1)(2T'_{A-1}+1)(2t+1)(2t'+1)} 
     \cr & & \hspace{2.5cm} \times
    \sixj{T_{A-2}}{\frac{1}{2}}{T_{A-1}}{\frac{1}{2}}{T_{A}}{t}
    \sixj{T_{A-2}}{\frac{1}{2}}{T'_{A-1}}{\frac{1}{2}}{T'_{A}}{t'} 
     \cr & & \hspace{1.5cm} \times
    \sum_{t_z} \bra T_{A-2} T_{(A-2)z} T_A T_{Az}   | t  t_z \ket
               \bra T_{A-2} T_{(A-2)z} T'_A T_{Az}  | t' t_z \ket
     \cr & & \hspace{3.5cm} \times
    \bra K_N  (\ell_N ;s )j \; t  t_z |\; H_2 \;| 
         K'_N (\ell'_N;s')j \; t' t_z \ket_{K_{N-1}} 
\end{eqnarray}
 where $S_n$ ($T_n$) is the total spin (isospin) of the particles 
1, 2,..., n. The recoupling leads to the following result for the matrix 
elements of $H_2$ depending on the A-body HH functions and on the spin-isospin 
states of the last two particles   
\begin{eqnarray} & & \hspace{1.5cm} \label{H2b}
    \bra K_N  (\ell_N ;s )j \; t t_z |\; H_2 \;| 
         K'_N (\ell'_N;s')j \; t' t_z \ket_{K_{N-1}} = 
  \cr & & \hspace{1cm}
   \delta_{K_N ,K'_N}\delta_{\ell_N,\ell'_N}\delta_{s,s'}\delta_{t,t'}
   \frac{1}{2 m} \frac{K_N(K_N+3N-2)}{\rho^2} 
  + V^{K_{N-1}\,j}
     _{K_N \ell_N s t t_z, K'_N \ell'_N s' t' t_z}(\rho) 
  \;,
\end{eqnarray}
where 
\begin{equation}
   V^{K_{N-1}\,j}
     _{K_N \ell_N s\, t\, t_z, K'_N \ell'_N s' t' t_z}(\rho) 
  = \sum_p W^{(p) \; K_{N-1}}_{K_N \ell_N, K'_N \ell'_N }(\rho)
           O^{(p)\;j}
            _{ \ell_N s\, t\,t_z, \ell'_N s' t' t_z} \; .
\end{equation}
Here 
\begin{equation}
W^{(p) \; K_{N-1}}_{K_N \ell_N, K'_N \ell'_N }(\rho) = 
\int d \Omega_N {\cal Y}_{[K_{N}]}^* 
                   V^{(p)}(\sqrt{2} \rho \sin \alpha_{N})
                   {\cal Y}_{[K_{N}']} \; ,
\end{equation}
and  
\begin{equation}
           O^{(p)\;j}
            _{ \ell_N s t t_z, \ell'_N s' t' t_z} =
  \bra (\ell_N ;s )j \; t t_z
  | \hat{O}^{(p)} | 
       (\ell'_N;s')j \; t' t_z \ket \;,
\end{equation}
is the usual 2-body matrix element for the operator $\hat{O}^{(p)}$.
One sees that $H_2$ is diagonal in the quantum numbers $[K_{N-1}],j, t_z$ 
and 

normally, as in case of the potential models considered in this work, also
diagonal in $s$ and $t$.

Due to the hyperangular integration 
$H_2$ explicitly depends on quantum numbers of the residual system, i.e.
$K_{N-1}$, while it is independent of the other quantum numbers in $[K_{N-1}]$.
As a result the  below defined  HH 
effective interaction depends on the state of the residual $A-2$ particle
subsystem.
 Such a "medium correction" of the 2-body force is of course a great 
advantage. On the other hand one has to pay for it with  greater numerical 
effort, since the effective interaction has to be calculated for all the 
various states and in addition it also depends on the specific $A$-body system 
considered. It is similar to the HO multi-valued effective 
interaction ~\cite{Zheng95,Navratil96a}.
We solve the hyperradial equation on a 
grid, where $H_2$ is diagonalized for each grid point $\rho_i$ and for all
the possible values of $K_{N-1},j,
s,
t,t_z$ in our model space. 

Having defined $H_2$ we employ in Ref.~\cite{BLO00} the Lee-Suzuki 
~\cite{Suzuki80} similarity transformation method to construct the effective
interaction. We follow an analogous 
procedure as that of Barrett and Navr\'atil ~\cite{Navratil98} for
the HO effective interaction. To this end the
eigenvectors, $\{|i\ket\}$, and eigenvalues, $\{\epsilon_i\}$, of 

$H_2(\rho)$, given by Eq.~\ref{H2b}, 
are used to construct the effective interaction.
Denoting by $|\alpha\ket$ the HH functions that belong to our model 
space, i.e. the HH function $|[K_N]\ket$ with $K_N \leq K_{max}$, and by 
$|\beta\ket$ the states that belong to the $Q$ space, 
$Q=\{ |[K_N]\ket\;;\; K_N > K_{max} \}$
the Lee-Suzuki effective Hamiltonian takes the form,
\begin{equation}
  P \tilde{H}_{2} P = P H_2 P + P H_2 Q \omega P \; , 
\end{equation}
where the transformation operator $\omega=Q \omega P$ is given by the
equation
\begin{equation}\label{omega}
  \bra \beta | i \ket = \sum_{\alpha} \bra \beta | \omega | 
\alpha \ket \bra \alpha | i\ket \;.
\end{equation} 
If $n_P$ is the number of model-space HH basis functions that belong to the 
subspace $[K_{N-1}]$, we may solve Eq. (\ref{omega}) for $\omega$ by choosing 
a set, $\cal{A}$, of $n_P$ eigenvectors with the lowest eigenvalues $|i\ket$ 
and inverting the matrix $\bra \alpha | i\ket$.
The resulting effective 2-body Hamiltonian 
\begin{equation}\label{H2tilde}
 \bra \alpha | \tilde{H_2}(K_{N-1},\rho) | \alpha'\ket = 
  \sum_i^{n_P} \left[ 
       \bra \alpha | i \ket \epsilon_i \bra i | \alpha' \ket +
       \sum_{\beta} \bra\alpha | i\ket \epsilon_i \bra i|\beta \ket
       \bra \beta | \omega | \alpha \ket \right]
   \;,
\end{equation} 
will have the property that $P|i\ket$, $|i\ket\in\cal{A}$, is a right 
eigenvector of 
$\tilde{H_2}$ with eigenvalue $\epsilon_i$.
The effective Hamiltonian is in general a non-hermitian operator, however
it can be hermitized, using the transformation ~\cite{Suzuki82}
\begin{equation}\label{H2eff}
  H_{2\;eff} = [P(1+\omega^{\dagger}\omega)P]^{1/2}\tilde{H_2}
               [P(1+\omega^{\dagger}\omega)P]^{-1/2} \;.
\end{equation}
The effective interaction can be now deduced from $H_{2\;eff}$, by 
subtracting the kinetic energy term,
\begin{equation}\label{Veff}
  V_{eff} = H_{2\;eff} - \frac{1}{2 m} \frac{\hat{K}_N^2}{\rho^2} \;.
\end{equation}

\section{Numerical results}
\label{sec:num}

We apply the EIHH method to the three- and four-body nuclei calculating ground 
state energies $(E_B)$, matter radii ($<r^2>^{1/2}$), and two-body correlation 
functions. We have chosen
to perform the calculation with two standard noncentral potential models, 
AV6 and AV14 \cite{AV14}. They are frequently used in other few-body approaches 
and thus allow to compare to other results in the literature. The model AV14 
is a fully realistic potential, while AV6 contains besides spin and isospin 
dependent central force only the tensor forces in addition.

In Fig.~1 we show the triton binding energy as function of the hyperspherical 
quantum number $K$. The convergence pattern looks very similar for both 
noncentral potential models. One sees that quite precise results are already 
obtained with rather small $K$ values. In fact deviations from the converged 
results amount to less than 0.1 MeV and 0.01 MeV for $K\ge 8$ and $K\ge 14$, 
respectively. It is evident from the figure that the additional potential terms 
of AV14 do not have any important impact on the convergence behavior.  In 
Fig.~1 we illustrate in addition our EIHH results from Ref.~\cite{BLO00} for 
the central potential model MTI-III \cite{MT}. The comparison with the AV6 
result shows that the tensor force does not lead to a deterioration of the 
convergence.

In Fig.~2 the results for the triton radius are shown. One sees that also
in this case an extremely good convergence behavior is obtained. For $K\ge 6$ 
one has only very small deviations of less than 0.005 fm from the converged AV6 
and AV14 results. As for the binding energy additional potential terms of AV14 
do not affect the convergence behavior. Again we 
show the MTI-III results. Contrary to the case of the binding energy the 
convergence is slowed down a bit by the tensor force.

Binding energy and radius are typical examples of observables which are not
very much affected by high momentum components of the wave function. One has to 
realize that our effective interaction approach can be considered in some way 
as a momentum expansion, since with increasing $K_{max}$ higher and higher
momentum components can be incorporated in the wave function. To study this
question better we show in Fig.~3 the two-body
correlation function
\begin{equation}
\rho(r) = \langle \Psi | \delta(r - r_{A,A-1}) | \Psi \rangle \,,
\end{equation}
where $\Psi$ is the ground state wave function ($\int dr \rho(r) = 1$).
One sees in Fig.~3a that $\rho(r)$ exhibits a rather broad and asymmetric 
maximum at about 2 fm. The convergence pattern of Fig.~3b shows that at larger
distances convergence is obtained with rather low values of $K$. Further
increase of $K$ improves the convergence at shorter and shorter distances.
In Fig.~3b we also illustrate a result with the bare interaction. For an
intermediate distance between about 1 and 3 fm one finds a slight
overestimation of $\rho(r)$, while at smaller and larger $r$ the correlation
function is considerably underestimated.

For the three-nucleon system it is possible to take into account rather high 
$K$ values in the calculation, in fact, e.g., for the highest value we 
considered ($K=32$) one has 374 hyperangular basis functions. Adding just one 
further nucleon leads to a considerable enhancement of the number of basis
functions. Therefore our $^4$He calculation is carried out only up to $K=18$,
where one already has 1975 hyperangular basis functions.
 
The results for the $^4$He binding energy and radius are depicted in Figs.~4
and 5, respectively. One notes that the convergence behavior is again very
good. On the other hand the results do not converge as rapidly as for the
three-nucleon case. In fact with $K=8$ one still lacks about 1 MeV binding,
while only about one tenth of this value is missing for the triton case at
the same $K$. For the convergence of the radius the differences are smaller
between the $^3$H and $^4$He results. For $K=6$ the $^4$He radius is 
overestimated by 0.015 fm, which is only about three times larger than  the 
corresponding overestimation for the triton case. However, also for $^4$He
one finds a rather good convergence for the higher $K$ values. 

From Figs.~4 and 5 it is evident that there are no important differences 
between the convergence patterns of the AV6 and AV14 cases showing again that 
the additional AV14 potential terms have a rather unimportant effect 
on the convergence behavior. Also for $^4$He we illustrate the results of
the central potential model MTI-III in the figures. They exhibit a little 
better convergence behavior than the potential models with tensor force.

In Fig.~6 we show the two-body correlation function of $^4$He. The convergence
pattern is not as good as for the triton case of Fig.~3. Even at larger
distances one needs a higher value of $K$ in order to reach the same precision.
The lower convergence rate can be attributed to the higher density of the 
alpha particle.

In Table I we compare our results for the three-nucleon system with other 
results in the literature. Inspecting the table one sees a very good agreement
among the various methods. It is evident that the binding energy can be 
calculated with a precision of about 0.01 MeV in most methods. For the alpha 
particle we obtain the following results: $E_B=25.61(4)$ MeV, 
$<r^2>^{1/2}=1.4739(4)$ fm (AV6, no Coulomb force) and $E_B=24.34(5)$ MeV, 
$<r^2>^{1/2}=1.5297(3)$ fm (AV14, with Coulomb force). The comparison 
with other results in the literature is not as good as for the triton case,
e.g., comparing with the SVM result ~\cite{Varga97} one finds a difference of 
0.2 MeV for the binding energy with AV6. One has similarly large differences 
comparing our $E_B$ for AV14 with that of Ref. ~\cite{Viviani}. It shows
that further investigations of the various groups 
~\cite{Navratil98,BLO00,Carlson,Varga97,Kievsky,Kamimura90,Nogga} 
have to be made in order to reach a better agreement. 
Such a bench mark study is currently under way.

\section{conclusion}
\label{sec:con}
In this work we have applied the recently developed hyperspherical effective 
interaction method to noncentral potential models including the realistic
NN force AV14. In this approach the two-body effective interaction depends
on the A-body hyperradius and on the state of the A-2 rest system explicitly.
Our results for the ground state properties of three- and four-nucleon
systems show that the method leads to an excellent convergence of the
hyperspherical expansion. The convergence is extremely fast in case
of triton, while it is a bit slower for the alpha particle. The difference
is explained by the higher density of the four-nucleon system, since our 
approach can be interpreted as a kind of momentum expansion.

In view of further applications it will be important to incorporate also 
three-nucleon forces in the EIHH method. Therefore we would like to mention 
that, similar to the HO approach ~\cite{Navratil98,NavKa}, the present 
formalism can be extended to derive an HH three-body effective interaction.

\vskip 2truecm
Acknowledgement: We thank the Institute for Nuclear Theory at the University
of Washington for its hospitality and the Department of Energy for partial
support during the completion of this work. This research is also supported
by the Italian ministry for scientific and technological research (MURST).

\begin{figure}
\caption{Triton binding energy as function of the hyperangular quantum number
$K$ with various NN potentials models: AV14 (a), AV6 (b), and MTI-III (c).
The value for the largest $K$ is indicated by a dashed line.} 
\end{figure}

\begin{figure}
\caption{Triton matter radius as function of the hyperangular quantum number
$K$ with various NN potentials models with notations as in Fig.~1.}
\end{figure}

\begin{figure}
\caption{EIHH results for the two-body correlation function of triton with 
the AV14 potential with various $K$ relative to the $K=26$ result;
also shown is the $K=26$ result with the bare interaction.}
\end{figure}

\begin{figure}
\caption{Alpha particle binding energy as function of the hyperangular quantum 
number $K$ with various NN potentials model with notations as in Fig.~1.}
\end{figure}

\begin{figure}
\caption{Alpha particle matter radius as function of the hyperangular quantum 
number $K$ with various NN potentials model with notations as in Fig.~1.}
\end{figure}

\begin{figure}
\caption{EIHH results for the two-body correlation function of $^4$He with 
the AV14 potential with various $K$ relative to the $K=16$ result.}
\end{figure}

\begin{table}
\caption{
         Comparison of binding energies ($E_B$) in [MeV] and root mean square
radii ($<r^2>^{1/2}$) in [fm] obtained with the present effective interaction 
method in the HH formalism (EIHH) with results of other methods.
For EIHH the number in parenthesis indicates the variance with respect to the 
result obtained with $K=K_{max}-2$. The quality of the convergence can be 
inferred from Figs.~1, 2}

\begin{center}
\begin{tabular}
{lccccc}
                  & & AV6 & & AV14\\ \hline
Nucleus&Method [Ref.]&$E_B$&$<r^2>^{1/2}$&$E_B$&$<r^2>^{1/2}$
\\ \hline
\\
$^3$H & EIHH                & 7.1602(3)   & 1.77475(5) & 7.6814(3) & 1.77615(3)
\\
  &Faddeev ~\cite{Chen}     & 7.15        &   --       &  7.670    &   --     
\\
  &GFMC ~\cite{Carlson}     & 7.22(0.12)  & 1.75(0.10) &     --    &  --
\\
  &VMC ~\cite{Arriaga}      & --          & --         &  7.53     &  --
\\
  &SVM ~\cite{Varga97}      & 7.15        & 1.76       &    --     &   --      
\\
 & HH ~\cite{Kievsky}      & --          &   --       &  7.684    &   -- 
\\
 & CHH  ~\cite{ELOT}        & --          &   --       &  7.69     &   --
\\ \hline\

%

\end{tabular}
\end{center}
\end{table}

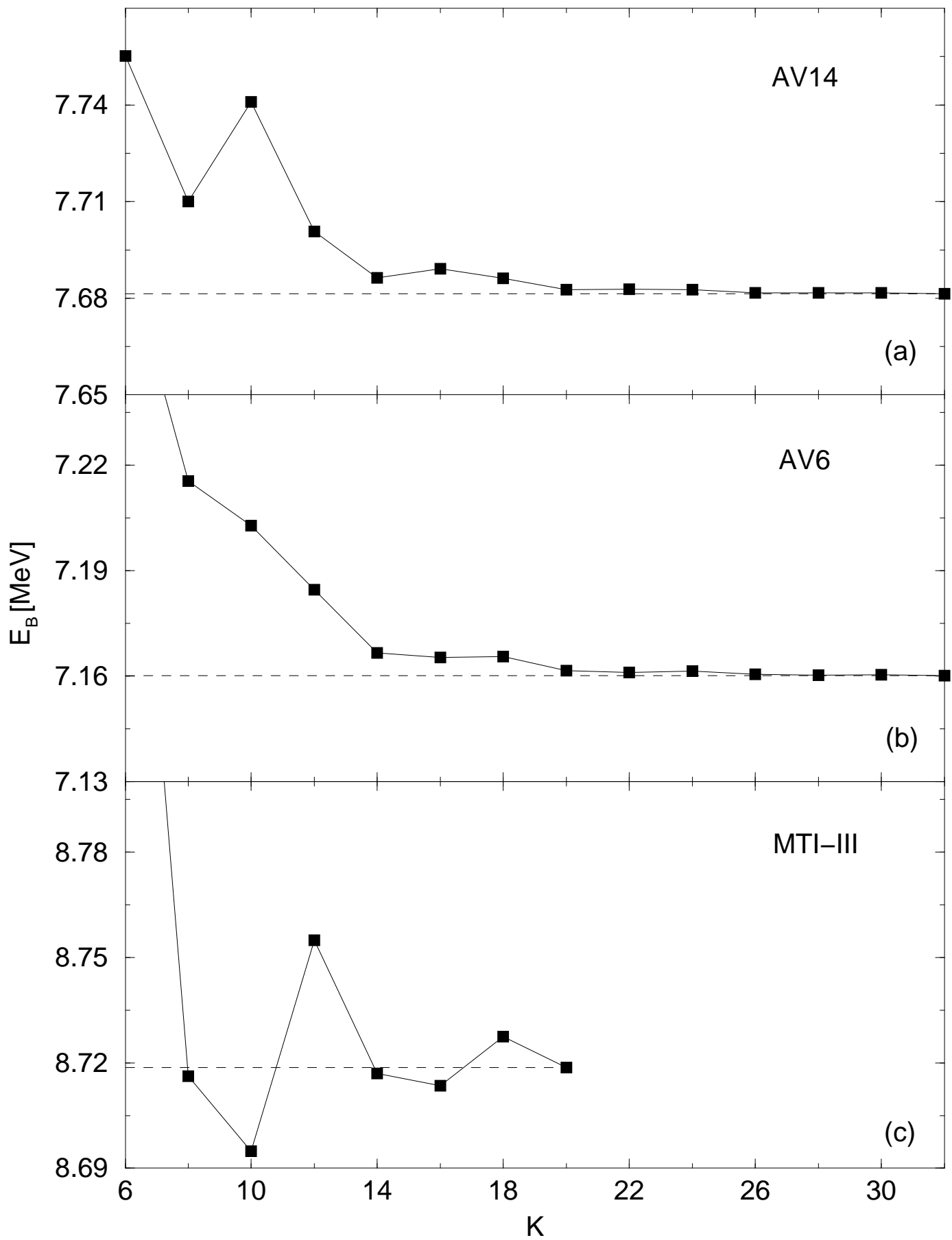
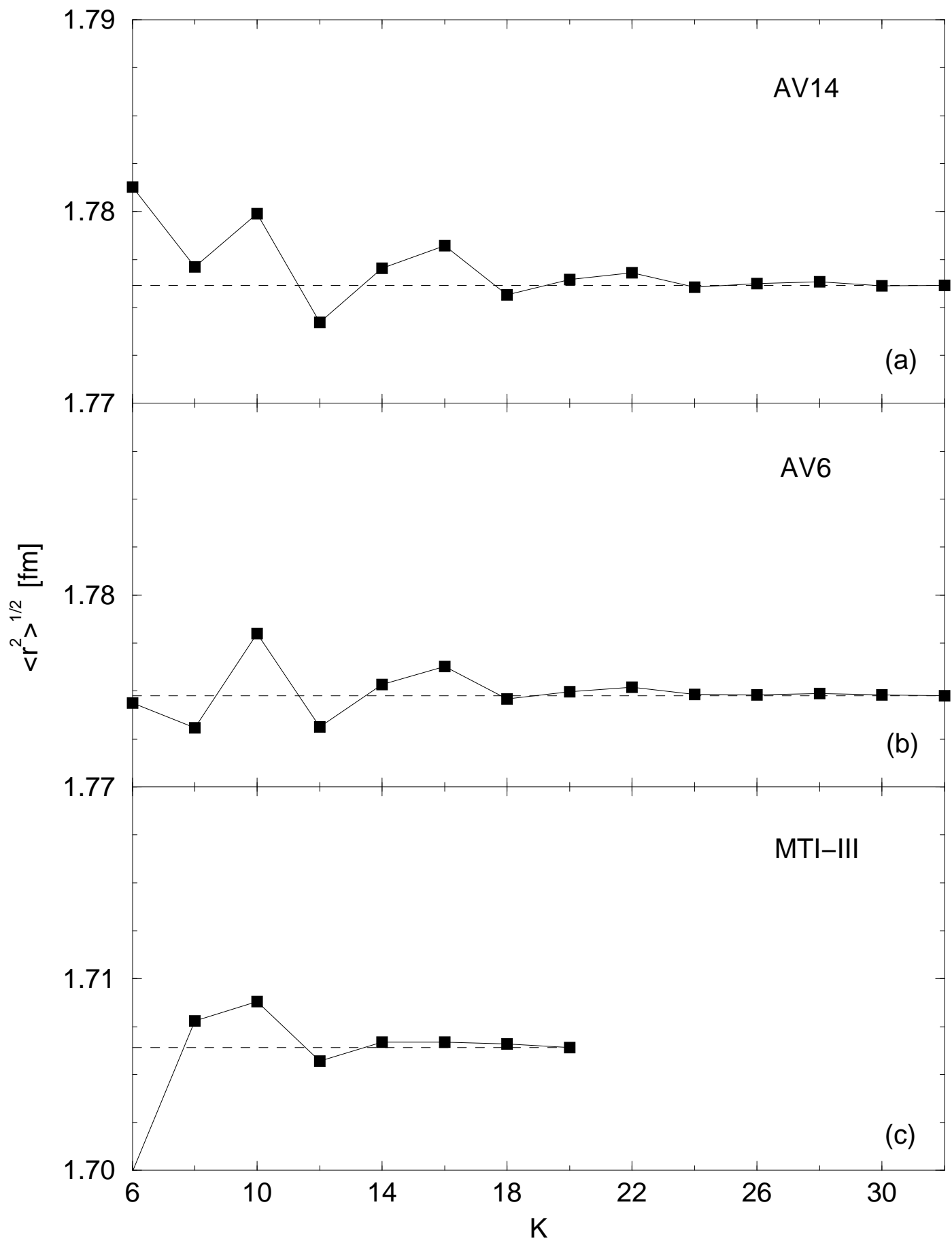
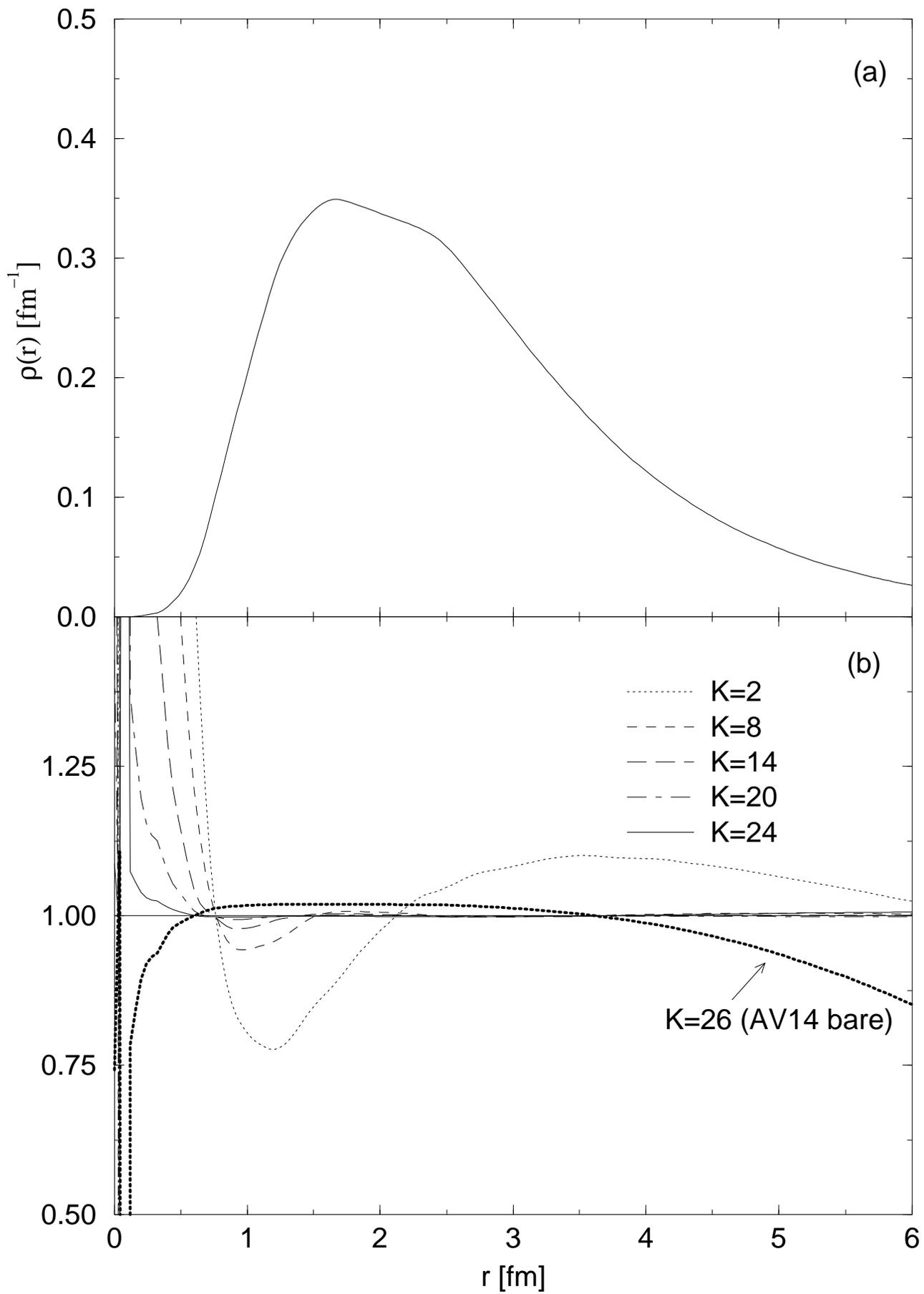
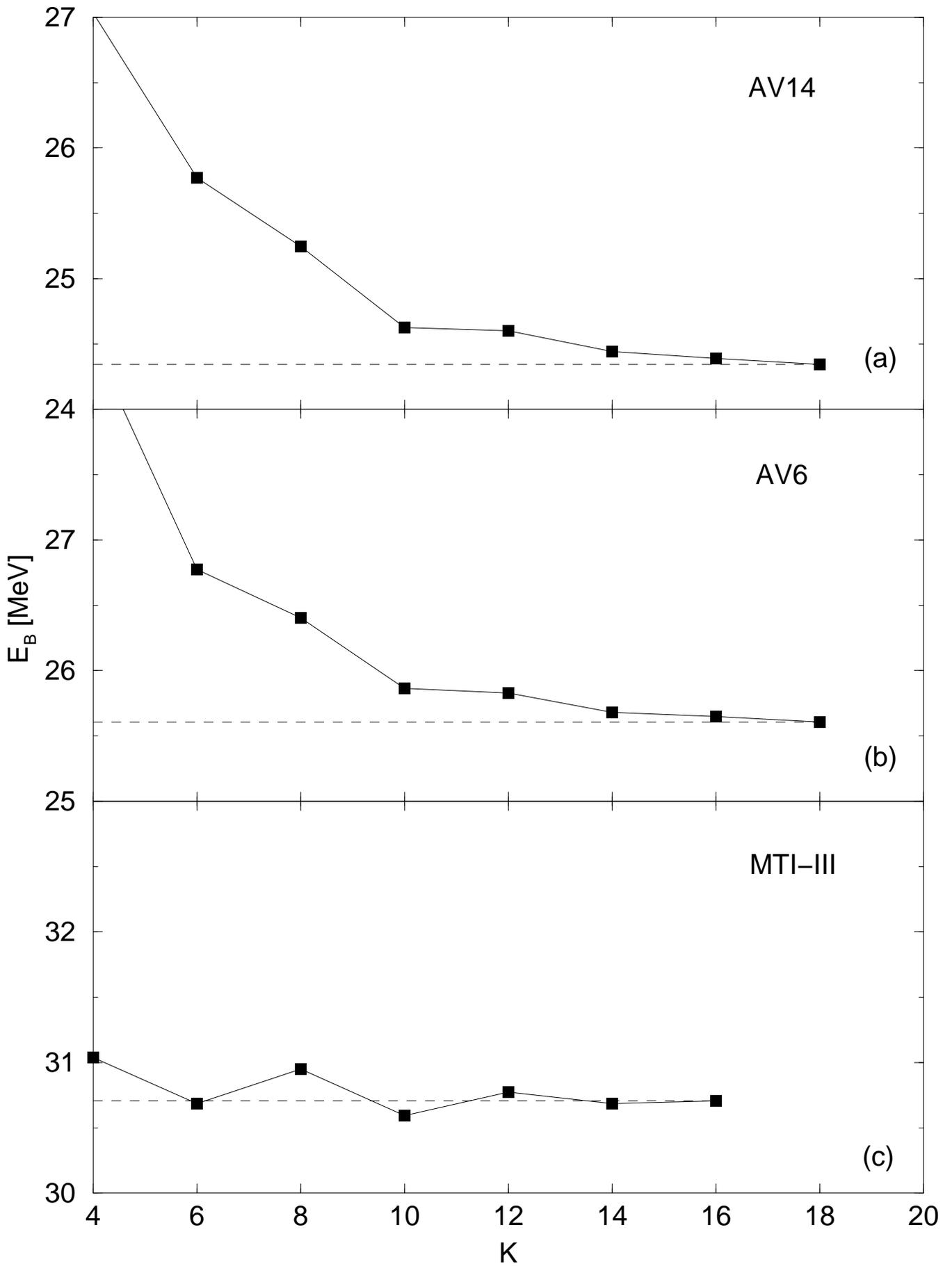
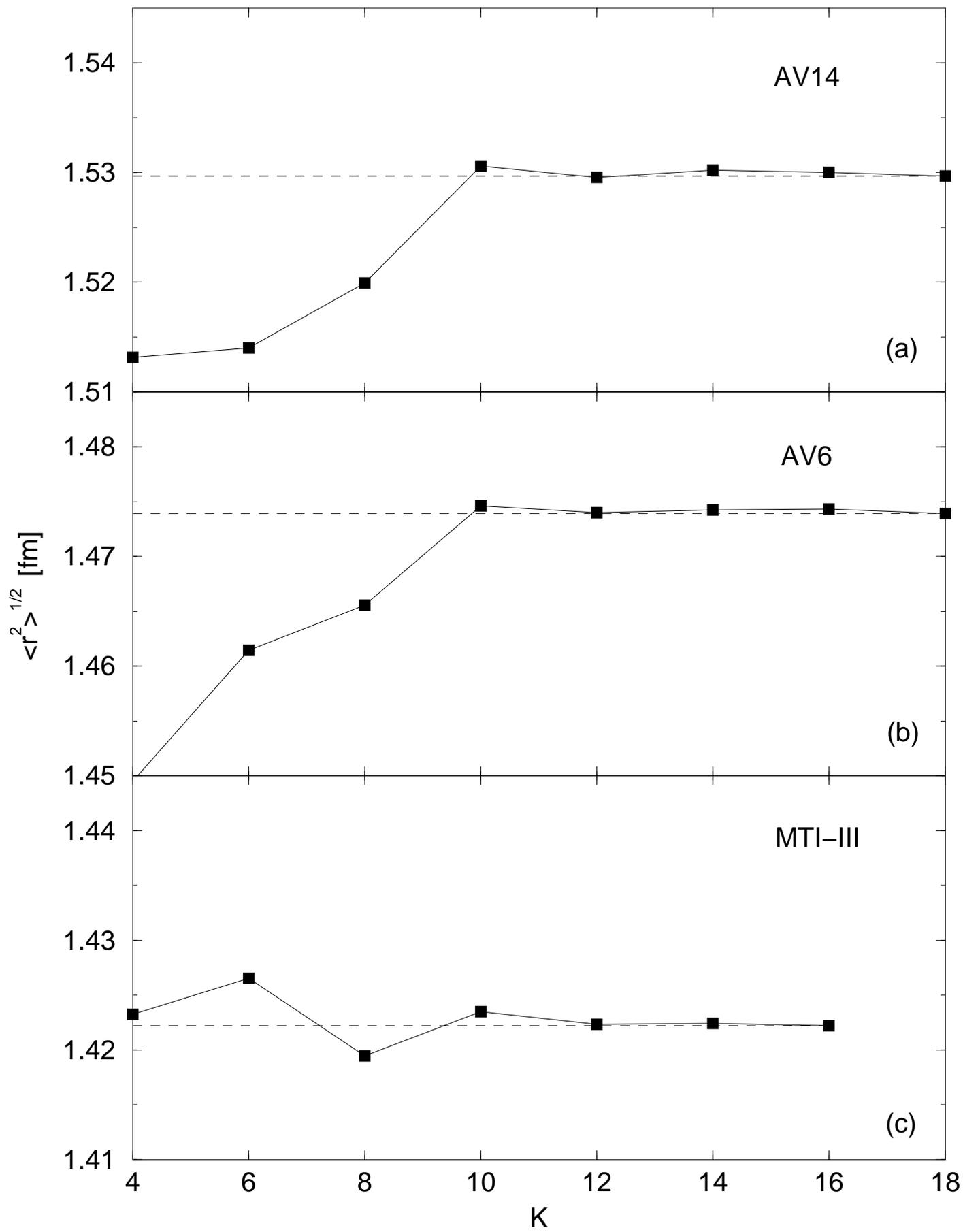
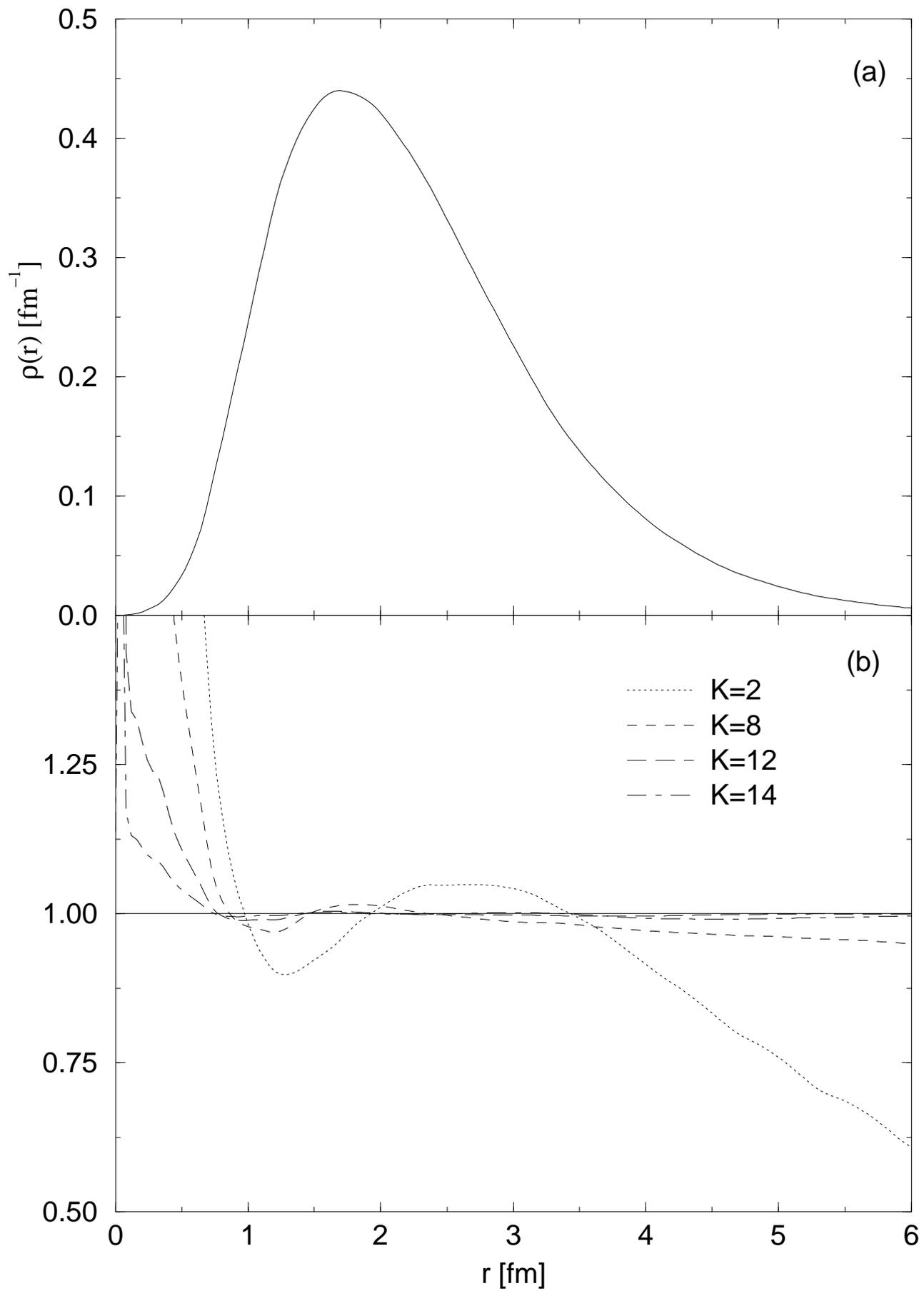
\end{document}